\RequirePackage{fix-cm}
\documentclass[smallextended]{svjour3}       % onecolumn (second format)
\smartqed  % flush right qed marks, e.g. at end of proof
\usepackage{graphicx}
\usepackage{multirow}
\usepackage{graphicx,subfigure}
\usepackage{subcaption}
\begin{document}

\title{Enhanced Tuberculosis Bacilli Detection using Attention-Residual U-Net and Ensemble Classification}


\author{Greeshma K.       \and
        Vishnukumar S. %etc.
}


\institute{Greeshma K. \at
              Department of Computer Applications, Cochin University of Science \& Technology, Kochi - 682 022, Kerala, India \\
              \email{greeshmavarier@gmail.com}            \\
             %\emph{Present address:} of F. Author  %  if needed
           \and
           Greeshma K. \at
              Department of Computer Science, Union Christian College, Aluva, Kochi - 683 102, Kerala, India \\
            %  \email{greeshmavarier@gmail.com}            \\
              \and
           Vishnukumar S. \at
              Department of Computer Applications, Cochin University of Science \& Technology, Kochi - 682 022, Kerala, India
}

\date{Received: date / Accepted: date}

\maketitle

\begin{abstract}
Tuberculosis (TB), caused by Mycobacterium tuberculosis, remains a critical global health issue, necessitating timely diagnosis and treatment. Current methods for detecting tuberculosis bacilli from bright field microscopic sputum smear images suffer from low automation, inadequate segmentation performance, and limited classification accuracy. This paper proposes an efficient hybrid approach that combines deep learning for segmentation and an ensemble model for classification. An enhanced U-Net model incorporating attention blocks and residual connections is introduced to precisely segment microscopic sputum smear images, facilitating the extraction of Regions of Interest (ROIs). These ROIs are subsequently classified using an ensemble classifier comprising Support Vector Machine (SVM), Random Forest, and Extreme Gradient Boost (XGBoost), resulting in an accurate identification of bacilli within the images. Experiments conducted on a newly created dataset, along with public datasets, demonstrate that the proposed model achieves superior segmentation performance, higher classification accuracy, and enhanced automation compared to existing methods. 
\end{abstract}

\section{Introduction}
Mycobacterium tuberculosis is the bacterium that causes tuberculosis (TB), a communicable bacterial infection. Although it usually affects the lungs, it can also damage the brain, kidneys, spine, or other organs. Tuberculosis can be fatal if left untreated. In the 2023 Global Tuberculosis Report \cite{whotbr2023}, tuberculosis was highlighted as the second leading cause of death globally from a single infectious agent in 2022, following COVID-19 and resulting in nearly double the number of deaths as HIV/AIDS. The report also indicated that a total of 1.3 million people succumbed to TB in 2022. Medical history, physical examinations, imaging tests such as chest X-rays, and laboratory testing such as sputum analysis or TB skin tests are frequently used in the diagnosis of tuberculosis. Not only can early detection of tuberculosis (TB) improve treatment outcomes, but it also prevents per capita transmission in the community. Presumptive examination is one of the methods for early detection. According to India TB report for 2022 \cite{tbr2022}, microscopy was used in 77\% (13,914,911) of the 18,035,463 presumptive exams that were conducted. The many benefits of microscopic examinations for tuberculosis (TB) detection include their affordability, non-invasiveness, accessibility, speedy outcomes, on-site diagnostic capabilities, and repeatability for precise evaluations.
\par 
Two types of microscopy, Fluorescent and Bright field, are employed for TB detection. To use bright field microscopy to detect bacteria, the Ziehl-Neelsen (ZN) staining method is used to prepare samples. At first thin smears of sputum which is collected from the patient  are prepared in a glass slide by fixing them and staining with carbol fuchsin. Under a microscope, acid-fast bacilli appear red or pink against blue or green, depending on the counterstain applied after the slide has been decolorized with acid-alcohol. The fluorescent microscopy needs Auramine - Rhodamine stained sputum samples prepared on glass slides. The sample smear is first stained with Auramine-O followed by rinsing to remove excess stain. A decolorizing agent, like acid-alcohol or acidified ethanol is applied then to remove non-acid-fast organisms and background staining. Finally, Rhodamine stain is applied to counterstain non-acid-fast bacteria and background material. In the slide thus prepared, the acid fast bacilli would appear as green color rods against black background. While Fluorescent microscopy boasts 10\% higher sensitivity than Bright field microscopy on average, the latter stands out for its cost-effectiveness \cite{steingart2006fluorescence}. 
In 2022, there were 7.5 million new TB diagnoses globally, officially recorded as TB cases, marking a rise above pre-COVID levels (7.1 million in 2019) by 16\%, surpassing 2021 levels by 28\%, and representing the highest annual count since the inception of global TB monitoring by World Health Organization (WHO) in the mid-1990s \cite{whotbr2022}. This surge in tuberculosis cases in 2022, adds to the workload of technical personnel, highlighting the need for an automated TB identification method. 
\par
Existing techniques in this field require upgrades in automation levels, aiming for more streamlined processes. Additionally,  improvements are needed in segmentation quality and classification accuracy to enhance the overall effectiveness of the methodologies. This paper presents an effective method for identifying tuberculosis bacilli, which integrates deep learning techniques for segmentation and employs an ensemble model based on machine learning for classification purposes. Section 2 describes the Related Works. Section 3 demonstrates the Methodology. Section 4 describes the Experimental Results and Analysis and section 5 concludes the paper.

\section{Related Works}
Attempts to automate the bacilli identification from bright field microscopic images were started in the late 2000s. The main aim of those works was to segment the bacilli region from the background. During that time, hand-crafted features were utilized to segment and quantify the bacterium. In \cite{costa2008automatic}, the analysis of images in the database included creating histograms with 10 gray level bands based on R-G values and establishing a threshold L through pixel quantity comparison. Size and morphological filters are used after it to eliminate large and small artifacts, respectively. The work reported a sensitivity rate of 76.65\% and a false positive rate of 12\%. The size and shape features along with the color feature to identify bacilli is proposed in \cite{sadaphal2008image}. Following an initial round of filtering using Bayesian segmentation, the size and the shape parameters, such as axis length ratio,eccentricity, and  area, were employed to identify bacilli. Other than designating the bacilli region, the results in the study were not quantified. In order to determine the degree of illness, the quantity of the bacteria can also be measured in addition to detecting the bacilli \cite{sotaquira2009detection}. Initially, the images are converted to the Lab and YCbCr color spaces. After the segmentation of each image independently, the results are combined using a logical AND operation. They claimed a sensitivity of 90.9\% and a specificity of 100\%.  The Average Bacilli Size (ABS) in pixels was found from the image dataset. Using this measurement the approximate number of bacilli is calculated from the segmented image irrespective of the bacilli presentations. In \cite{priya2015separation}, the method of concavity (MOC) outperforms traditional techniques like marker-controlled watershed (MCW) and multi-phase active contour (MAC) for the purpose of separating overlapping bacilli, with the MOC bacilli matching closest to true bacilli. 

\par 
Machine learning methods for identifying bacteria in images began around the same time as conventional image processing techniques, starting in 2009. In the realm of bacilli detection using machine learning, a common approach involves a two-step process comprising segmentation and classification. Segmentation identifies the region of interest (ROI), subsequently classified by the classifier into bacilli and non-bacilli categories. This method effectively minimizes the computational demands on the classifier by focusing its analysis solely on the relevant ROI rather than the entire image. R. Khutlang \textit{et.al.} used two one class classifiers to automate the bacilli detection process \cite{khutlang2010automated}. First, a pixel classifier was utilized to perform the segmentation, and then a second classifier was employed to determine the final bacilli regions. In \cite{khutlang2009classification}, the authors reported 88.38\% of correctly categorized pixels using the product of three pixel classifiers for segmentation. Five classifiers were tested during the classification phase. Using Fisher mapped features, all five classifiers provided greater than 95\% accuracy, sensitivity, and specificity during the classification step. A segmentation approach in two stages is proposed in \cite{zhai2010automatic}, combining the HSV and CIE Lab* color spaces. After segmentation, a classification algorithm uses three shape feature descriptors — area, compactness, and roughness — and employs a decision tree for judgment. Costa Filho \textit{et.al} \cite{costafilho2012mycobacterium} introduced a novel method, employing a feedforward neural network for image segmentation. The features were meticulously selected through a process of exploring various feature combinations. Additionally, they introduced a new feature called color ratio, which is also utilized in the classification process. Several handcrafted features were examined for segmentation, as in the case of \cite{costa2015automatic}, and it was found that R-G features produce the best segmentation results. However, they chose the top 4 features for the post-processing stage, where SVM \cite {hearst1998support} and feed forward neural networks were employed in addition to earlier techniques and some filters. SVM classifier yielded the best sensitivity rate result, which was 96.8\%. A random forest \cite{breiman2001random} based segmentation and classification was performed in \cite{ayas2014random} which gave better results than different machine learning models both in segmentation and classification phase. A Gaussian Mixture Model based segmentation followed by bacilli counting algorithm is used in \cite{fandriyanto2021detecting}. It provides an accuracy of 93.52\% where randomly chosen 8 images from the ZNSM-iDB database \cite{shah2017ziehl} are used for testing.

\par 
Among the studies that were used for identifying bacilli from bright field microscopic sputum smear images, Convolution Neural Network(CNN) is the dominant model used. Yadini Pérez López et al. \cite{lopez2017automatic} developed a three convolutional layer model that uses the image's R-G feature as input and reported a 99\% area under the ROC curve. A  deep neural network was used in \cite{kant2018towards} for TB detection which gives the positions of bacilli suspecting areas in the image. Though most of the authors use accuracy as a main evaluation metric, they emphasized precision (67.55 \%) and recall (83.78 \%) in their work. R. O. Panicker \textit{et.al} \cite{panicker2018automatic} proposed another CNN method, where the segmentation was carried out by the Otsu method. The Region of Interests(RoIs) were identified and fed into the CNN model for binary classification using connected component analysis. The model was evaluated by three metrics, recall (97.13\%), precision (78.4\%) and  F-score(86.76\%). M. El-Melegy \cite{el2019identification} applied the Faster RCNN for the identification of tuberculosis bacilli followed by a CNN to reduce the number of false positives. Accuracy is not reported in the study where as the Recall, Precision and F-score are 98.4\% 85.1\% and 91.2\% respectively. The model proposed by Dinesh Jackson Samuel \textit{et.al}\cite{dinesh2019tuberculosis}, produces an accuracy of 95.05\% which uses a transfer learning method based on Inception V3 DeepNet model and SVM classifier. Two CNN models are used in \cite{yang2020cnn} for forming a pipeline along with the Logistic Regression (LR) model.It yields a sensitivity of 87.13\%, specificity of 87.62\% and F1 score of  80.18\%. A novel idea of making a mosaic image to detect bacilli from the original microscopic image was introduced by M. K. M. Serrao \textit{et.al}  in \cite{serrao2020automatic}. It requires more effort to make a mosaic image dataset by avoiding clusters and fragments which makes it less applicable in the real world scenarios although the accuracy is 99\%. R. O. Panicker \textit{et.al}, proposed a lightweight CNN model in \cite{panicker2021lightweight} with accuracy 97.83\% with a lesser number of parameters 18,213. Later in 2022, R. O. Panicker \textit{et.al} \cite{panicker2022automatic} made a new Densely Connected Convolutional Networks using DenseNet-121 architecture which improved the classification accuracy to 99.7 \%. In \cite{greeshma2023identification} the images are initially segmented using a U-Net model, followed by classification with a Random Forest classifier, achieving an accuracy of 93.98\%. A recent update has been done by K.S., Mithra \cite{mithra2023enhanced} which used Otsu for segmentation and an enhanced Fuzzy Gaussian Network for classification with a segmentation accuracy of 92.4 \% and a MSE of 0.004\%.

\par 
Most of the existing works have directly fed the Regions of Interest (RoIs) into the model without employing an efficient segmentation model. In \cite{mithra2023enhanced} traditional Otsu segmentation algorithm is followed prior to the classification phase, however it struggles to adequately segment the Regions of Interest (RoIs). Segmentation is an inevitable step on the path to automating the identification of bacilli areas from microscopic sputum smear images. An efficient segmentation model will also help reduce the burden on the classifier by narrowing down the huge dataset into several small RoIs. In order to address this, the proposed approach utilizes an Attention Residual U-Net to efficiently extract Regions of Interest (RoIs). Most importantly, the efficacy of classifiers plays a pivotal role in shaping the overall performance of the model designed for tuberculosis bacilli detection from sputum smear images. There is still room for improvement for the classifiers used in the existing techniques. Ensemble classifiers combine the advantages of several individual classifiers, leveraging their collective strengths to enhance overall performance and accuracy in predictions. An ensemble classifier is proposed, which combines SVM, Random Forest and XGBoost classifiers, enhancing the accuracy and effectiveness of the classification process.

\par
Several advanced segmentation architectures have been proposed in recent years to address specific challenges in medical image segmentation. OFF-eNET introduces a fully end-to-end 3D CNN utilizing dilated convolutions and residual connections to achieve precise segmentation of thin intracranial blood vessels \cite{nazir2020off}. NHBS-Net employs a dual attention mechanism and feature fusion to enhance segmentation accuracy in ultrasound imaging, particularly in detecting key anatomical structures \cite{liu2021nhbs}. Similarly, ECSU-Net utilizes a sliced U-Net architecture combined with clustering and fusion strategies for efficient segmentation and classification of vertebrae from 3D medical images \cite{nazir2021ecsu}.Advanced medical image segmentation architectures have demonstrated significant progress in addressing challenges such as feature extraction, multi-scale processing, and contextual understanding. The CaVMamba model integrates convolution and VMamba architectures, employing a dynamic feature fusion module to enhance segmentation by effectively combining local and global features \cite{chen2024cavmamba}.Similarly, MFADU-Net leverages a multi-level feature fusion block and an adaptive atrous convolution decoder to refine segmentation performance, particularly for complex anatomical boundaries and multi-scale features \cite{zhao2024mfadu}. These developments highlight the growing trend of integrating hybrid architectures to improve the accuracy and robustness of medical image segmentation tasks.

\par 
The foundation of existing works heavily relies on the utilization of public datasets \cite{ya12-j913-22} or \cite{shah2017ziehl}. Using public datasets exclusively in identification of bacilli may potentially limit their generalization to new data. Introducing a new dataset helps overcome this risk by providing diverse data for better  generalization and enhanced performance analysis. We have created a new dataset 'DCA-CUSAT Bright Field Microscopic Sputum Smear TB Dataset' comprising 101 bright field microscopic sputum smear images captured from TB-positive ZN stained slides obtained from Government District TB Hospital, Ernakulam, Kerala, India.

\section{Methodology}
 The proposed approach consists of two phases, segmentation and classification. Initially, the images in the dataset are segmented using QuPath \cite{bankhead2017qupath} to obtain the binary masks, which are then used to train and evaluate both segmentation and classification models. For segmentation, a Residual U-Net with Attention gates attached in the skip connections is used.  An ensemble classifier is used in the proposed method for classification which combines  three classifiers, SVM \cite{hearst1998support}, Random Forest \cite{breiman2001random} and XGBoost \cite{chen2016xgboost}.

\begin{figure}[htp]
    \centering
    \includegraphics[scale=0.1]{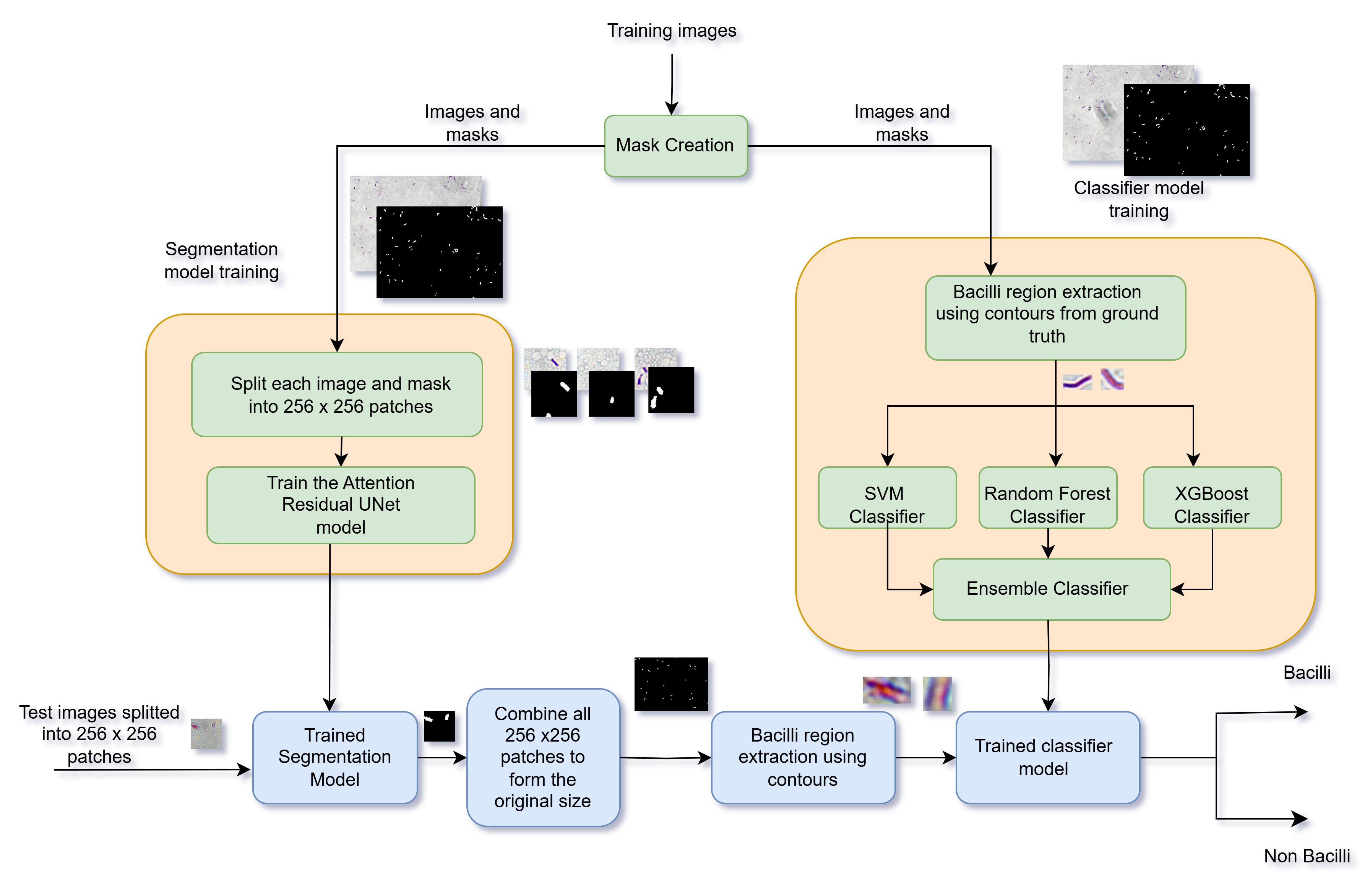}
        \caption{Proposed System Architecture.}
    \label{oas}
\end{figure}

\begin{figure}[htp]
    \centering
    \includegraphics[scale=0.15]{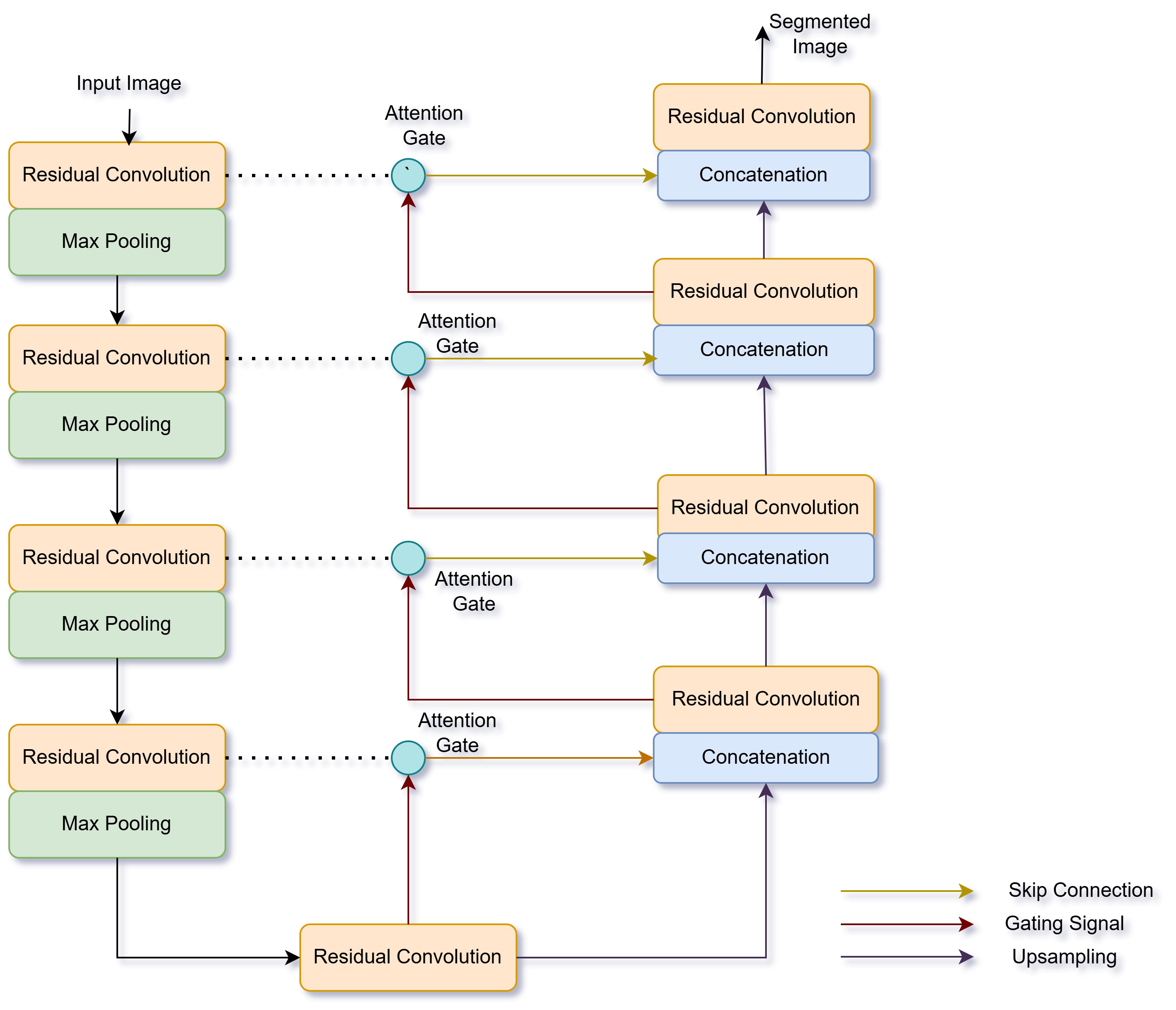}
    \caption{Proposed Attention Residual U-Net for segmentation.}
    \label{pau}
\end{figure}

The overall architecture of the proposed model is shown in Figure 
\ref{oas}. The performance of segmentation using the U-Net model is significantly enhanced by incorporating residual connections and attention mechanisms\cite{alom2018recurrent}, \cite{oktay2018attention}. The proposed method utilizes the U-Net model with residual connections in the encoder path and attention mechanisms in the skip connections. The architecture of the proposed Attention Residual U-Net  is  shown in Figure \ref{pau}.

\par 
The model begins with downsampling layers that reduce the spatial dimensions of the input features while increasing the number of filters to extract hierarchical features. Each downsampling step includes a residual convolutional block that consists of convolutional layers followed by batch normalization, dropout, and ReLU activation. These blocks help the model to learn rich feature representations at multiple scales.
\par
\begin{figure}[htp]
    \centering
    \includegraphics[scale=0.13]{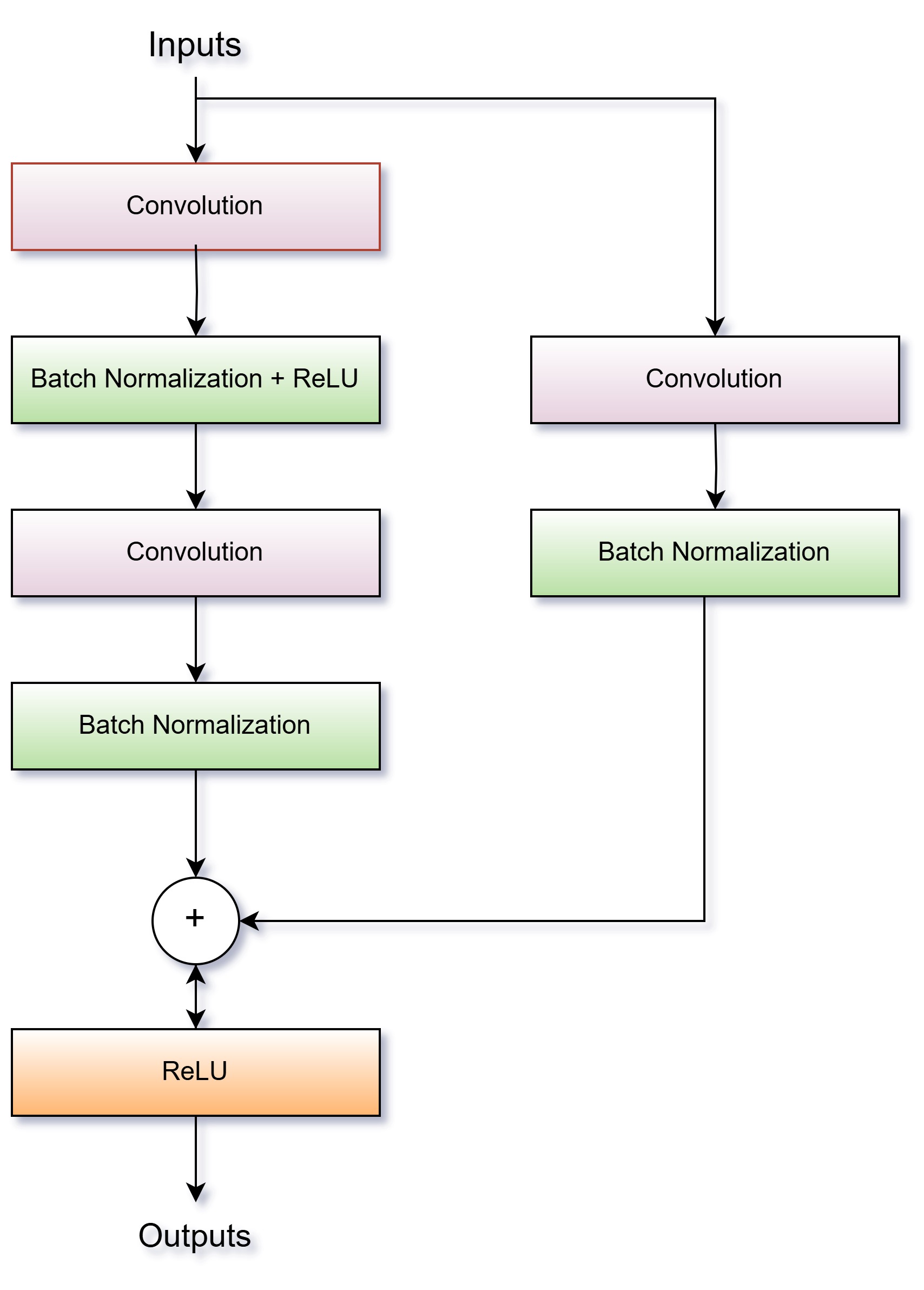}
    \caption{Architecture of the Residual Convolution.}
    \label{arc}
\end{figure}

 
In the upsampling layers, attention mechanisms are introduced to selectively emphasize important features while suppressing irrelevant information. The gating signal generation function creates a gating feature map that is used to modulate the importance of features at different spatial locations. This attention mechanism allows the model to focus on relevant regions, improving segmentation accuracy. Additionally, the model employs residual connections that enable the flow of information across different layers, aiding in the efficient propagation of gradients during training.  The architecture of the residual convolution is depicted in Figure \ref{arc}.

\par 
In the classification stage of the proposed methodology, an ensemble classifier, that combines three classifiers, SVM, Random Forest and XGBoost is employed, taking advantage of their learning strategies. The ensemble approach allows each classifier to specialize and learn unique patterns and features within its designated subset, enhancing their understanding of the dataset's complexities. 
\par
The SVM classifier \cite{hearst1998support} in the ensemble classifier, uses the Radial Basis Function (RBF) kernel for modeling non-linear decision boundaries.  It works well in situations where data points may not be easily separable by a linear boundary and can effectively model non-linear relationships. Its ability to find the optimal hyperplane that maximizes the margin between classes makes it a valuable tool for classification tasks.
\par
Random Forest \cite{breiman2001random} is a versatile ensemble learning method well-suited for image classification tasks due to its capability to handle high-dimensional image data and capture complex interactions between image features. This method constructs multiple decision trees during training and combines their predictions to enhance accuracy making it effective in distinguishing between different image classes.
\par
XGBoost \cite{chen2016xgboost} is a robust machine learning algorithm known for its ability to leverage weak learners, in a sequential manner. In image classification tasks, XGBoost's iterative approach of adding weak learners and focusing on minimizing errors allows it to capture complex patterns and relationships within image data effectively. Its computational efficiency and speed make it a popular choice for achieving accurate results across diverse datasets.
\par
Initially the above three classifiers are trained separately using a disjoint subset of training data. Following individual training, the three classifiers were combined into an ensemble classifier. The ensemble classifier was further trained using all three subsets of the training data. In the ensemble classifier, the predictions from all three base classifiers were combined using the hard voting approach. In hard voting, the class that receives the most votes among the base classifiers is chosen as the predicted class for a given input instance. This final training phase facilitated the amalgamation of insights and expertise from the individual classifiers, resulting in an ensemble model that excelled in classifying bacilli regions accurately. 

\section{Experimental Results and Analysis}
For the experiments, we utilized a powerful computing facility located at the Department of Computer Applications, Cochin University of Science and Technology in Kochi, Kerala, India. The setup includes 2 NVIDIA A100 40GB PCIe 4.0 GPU cards, each with 6912 CUDA cores and 432 Tensor cores. These GPUs are supported by 2 Xeon Gold 6226R 16C 150W 2.9GHz processors and 12 $\times$ 64 GB DDR4 2933 MHz RDIMM memory modules, providing substantial computational capabilities for deep learning and other intensive tasks. Additionally, the system is equipped with 8 $\times$ 1.92 TB 6Gbps SATA 2.5'' SSDs for fast data storage and retrieval. 
\par
The experimental work comprised three main phases, the generation of a new dataset, the evaluation of the proposed model's performance on both new dataset and two public datasets, and a comparative analysis to evaluate the performance of the proposed method against recent existing methods using aforementioned datasets.

The performance of the proposed model is evaluated in both segmentation and classification stages. The segmentation model is evaluated using two metrics, Jaccard Index and Dice coefficient. Jaccard Index or Intersection over Union (IoU) measures the proportion of the intersection of the segmented regions to the union of these regions, of the predicted mask and the ground truth mask. The Jaccard Index, $J$ is defined as

\begin{equation}
    J = \frac{A \cap B}{A \cup B}
\end{equation}

where $A \cap B$ represents the number of pixels that are correctly segmented in both the predicted mask (A) and the ground truth mask (B) and $A \cup B$ represents the total number of pixels segmented either in the predicted mask or in the ground truth mask.

Dice coefficient measures the ratio of the intersection of the segmented regions to the average size of the segmented regions in the predicted mask and in the ground truth mask. The Dice Coefficient, D is defined as

\begin{equation}
    D = 2 \times \frac{|A \cap B|}{|A| + |B|}
\end{equation}

Where $|A \cap B|$ represents the number of pixels that are correctly segmented in both the predicted mask (A) and the ground truth mask (B) and $|A|$ and $|B|$ denote the total number of pixels segmented in the predicted mask and ground truth mask, respectively.

The performance of classifiers is evaluated based on their accuracy, precision, recall and F1 score. For every class, the properly recognised prediction is called True Positive (TP), the correctly rejected prediction is called True Negative (TN), the mistakenly identified forecasts are called False Positive (FP), and the wrongly rejected predictions are called False Negative (FN).
The accuracy, precision, recall and F1 score are calculated using equations \ref{acc_equ}, \ref{pre}, \ref{re} and \ref{f1} respectively.

\begin{equation}
    Accuracy = \frac{TP + TN}{TP + TN + FP + FN}
    \label{acc_equ}
\end{equation}

\begin{equation}
Precision = \frac{TP}{TP + FP}
\label{pre}
\end{equation}

\begin{equation}
    Recall = \frac{TP}{TP+FN}
    \label{re}
\end{equation}

\begin{equation}
    F1~Score = 2 \times \frac{Precision \times Recall}{Precision + Recall}
    \label{f1}
\end{equation}

\par

In the literature, majority of tuberculosis bacilli detection methods from bright filed microscopic sputum smear images exclude the segmentation phase to identify RoIs. However, two recent methods \cite{panicker2018automatic} and \cite{mithra2023enhanced} have employed segmentation stages for detecting Regions of Interest (RoIs). We compare the performance of the proposed method against these two methods to validate the superiority of the proposed method.

\subsection{DCA-CUSAT Bright Field Microscopic Sputum Smear TB Dataset}
\begin{figure}[htp]
    \centering
    \includegraphics[scale=0.4]{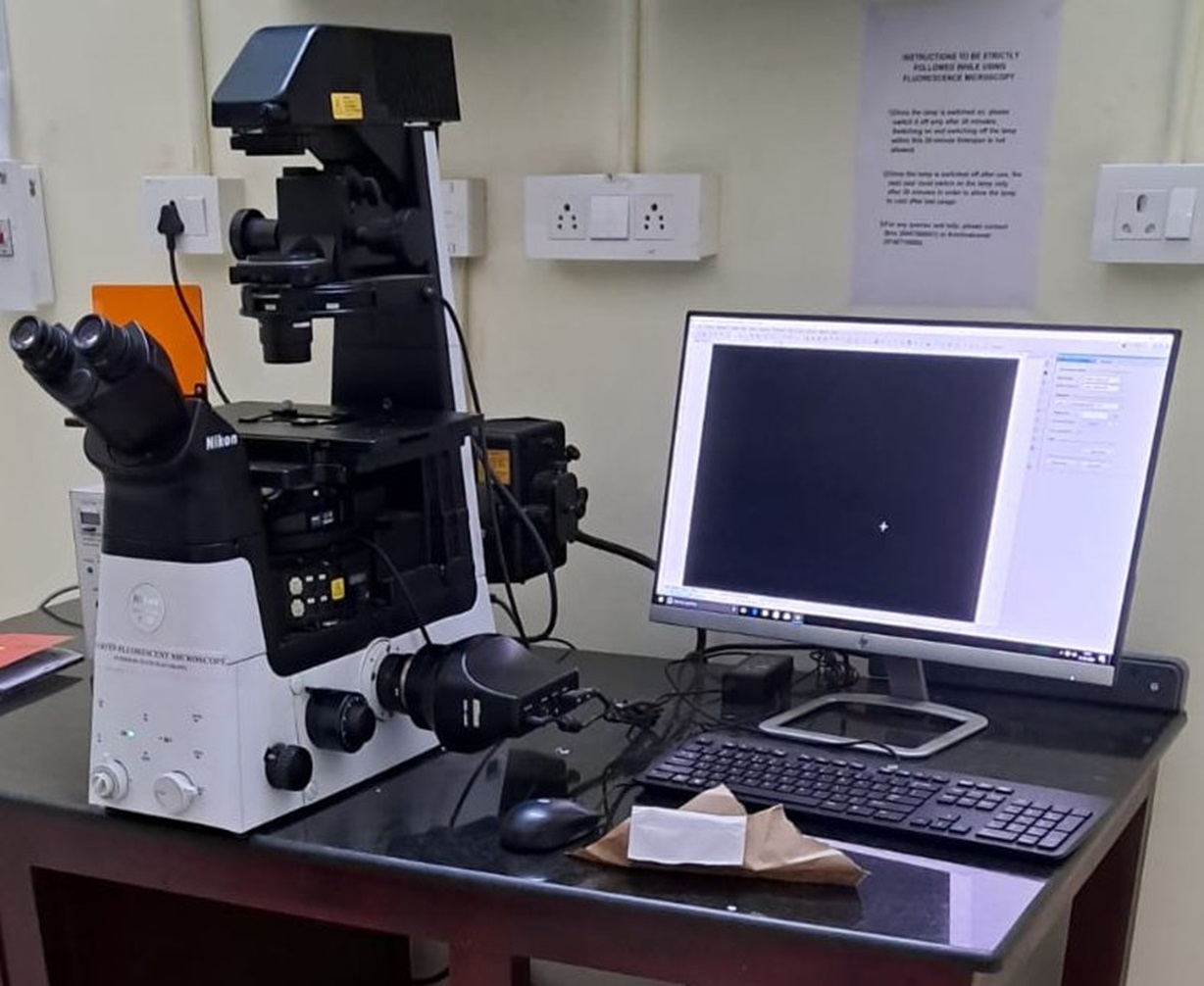}
    \caption{Nikon Ti2-u Eclipse microscope, integrated with the NIS-elements software package at Department of Biotechnology, Cochin University of Science and Technology. }
    \label{mi}
\end{figure}

For this research, we have curated a database comprising 101 microscopic images sourced from three Ziehl-Neelsen (ZN) stained sputum smear slides each containing sputum smears from patients with Tuberculosis infection rated at level 3+. The database has been named as 'DCA-CUSAT Bright Field Microscopic Sputum Smear TB Dataset' (DCA-CUSAT TB dataset). The ZN stained sputum smear slides were obtained from the Government District Tuberculosis Hospital, Ernakulam, Kerala, India, while the imaging process took place at the Microscope facility within the Department of Biotechnology at Cochin University of Science and Technology, Kochi, Kerala, India. The images were captured using the Nikon Ti2-u Eclipse microscope, integrated with the NIS-elements software package. The Nikon Ti2-u Eclipse microscope is specifically designed for a range of scientific and research applications, featuring advanced imaging capabilities that enable connectivity to a computer for image viewing and capture. This camera integrated system is shown in Figure \ref{mi}.

After applying a 50\% glycerol solution and placing a coverslip on the ZN stained sputum smear samples, images were captured at 100 $\times$ magnification using Nikon Ti2-u Eclipse microscope. The purpose of applying glycerol 50\% was the following:

\begin{enumerate}
    \item To preserve the sample by preventing dehydration and maintaining their structural integrity over time.
    \item Since the refractive index of glycerol is close to that of glass coverslips and immersion oil, it helps to reduce light scattering and improve the clarity and resolution of microscopic images.
\end{enumerate}

\begin{figure}[htp]
  \begin{center}
    \subfigure[Slide 1]{\label{sam1}\includegraphics[scale=0.3]{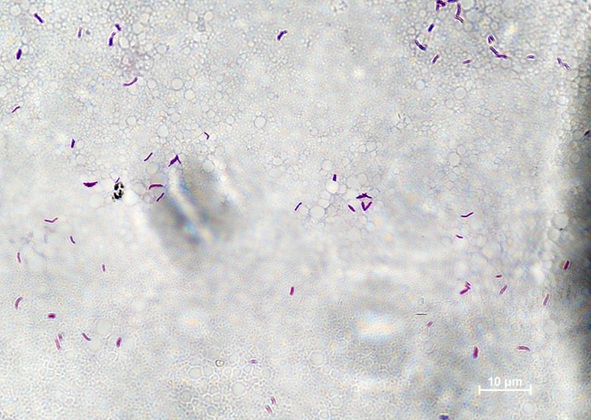}}
    \subfigure[Slide 2]{\label{sam2}\includegraphics[scale=0.3]{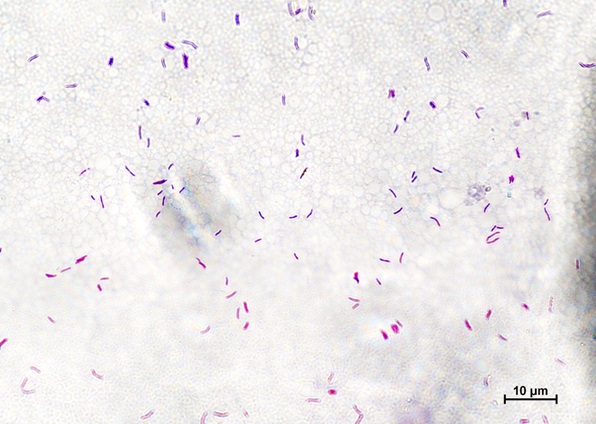}} 
    \subfigure[Slide 3]{\label{sam3}\includegraphics[scale=0.3]{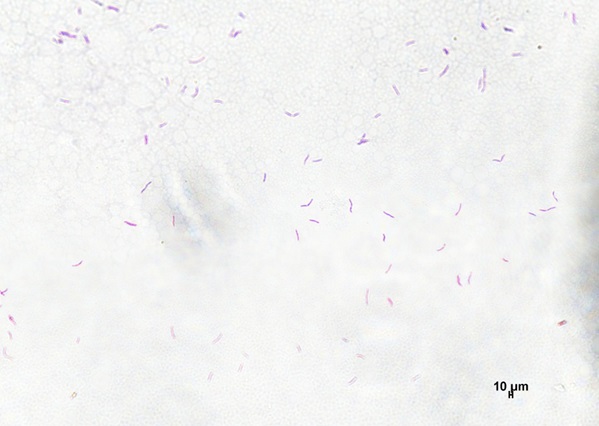}}
  \end{center}
  \caption{Sample images - DCA-CUSAT Bright Field Microscopic Sputum Smear TB Dataset.}
  \label{si}
\end{figure}

The images were captured by a linear pattern systematically moving across the specimen in a straight line from top left to bottom right. 31 images were taken from the first slide, and 35 images each were captured from both the second and third slides. All images maintained a resolution of 2880 $\times$ 2048 pixels. Sample images from each of the three slides are shown in Figure \ref{si}.

\subsection{Performance Analysis of the Proposed Model Using DCA-CUSAT TB dataset}

The evaluations of the Attention Residual U-Net, employed for segmentation, and the ensemble classifier, utilized for classification, were carried out using the DCA-CUSAT TB dataset. Among the 101 images, 81 were designated for training the proposed model, while the remaining 20 were reserved for testing purposes.

\begin{figure}[htp]
  \begin{center}
    \subfigure[Training accuracy]{\label{sam1}\includegraphics[scale=0.8]{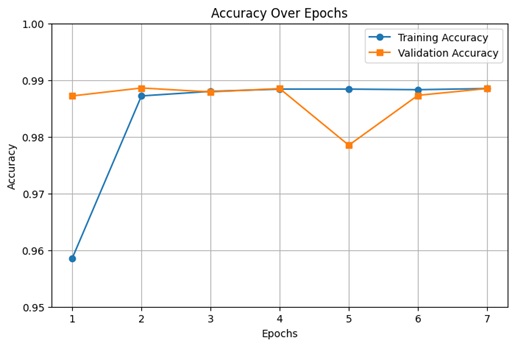}}
    \subfigure[Jaccard Index]{\label{sam2}\includegraphics[scale=0.6]{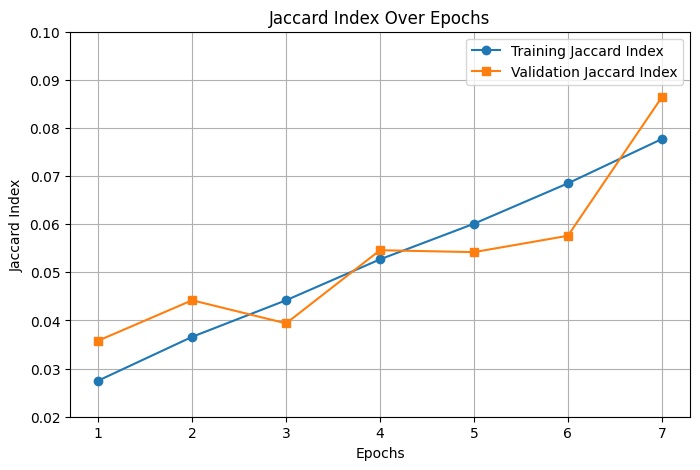}} 
    \subfigure[Loss]{\label{sam3}\includegraphics[scale=0.8]{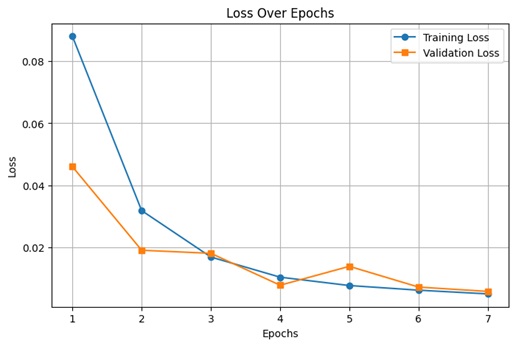}}
  \end{center}
  \caption{The progress of Attention Residual U-Net model training for segmentation on DCA-CUSAT TB dataset.}
  \label{gr}
\end{figure}

\par 
For segmentation, the training images and their respective masks were divided in both horizontal and vertical directions, into multiple patches having resolution 256 $\times$ 256. This process resulted in the creation of 7128 patches, each with its corresponding binary mask, which were utilized to train the Attention Residual U-Net model for segmentation, achieving an accuracy of 98.85\%. The training procedure was set for 20 epochs, incorporating an early stopping mechanism to avoid overfitting. The training concluded after the $7^{th}$ epoch due to consistent validation accuracy over the past consecutive epochs. The improvement in the values of training accuracy, jaccard index and loss are measured and plotted in Figure \ref{gr}.

\par
For training each sub-classifier(Random Forest, XGBoost, and SVM) within the ensemble classifier, 250 regions containing bacilli and 250 regions without bacilli were extracted from the images across the training set. The ensemble classifier, built from three sub-classifiers, was trained using a combined dataset of 750 bacilli regions and 750 non-bacilli regions. Table \ref{tr_acc} displays the accuracy of each sub classifier and the ensemble classifier on DCA-CUSAT TB dataset, during the training phase.

\begin{table}[htp]
\caption{Training Accuracy of individual classifiers and ensemble classifier}
\label{tr_acc}
\centering
\begin{tabular}{|c|c|c|c|}
\hline
\textbf{Classifier}    & \textbf{\begin{tabular}[c]{@{}c@{}}Number of \\ \\ bacilli regions\end{tabular}} & \textbf{\begin{tabular}[c]{@{}c@{}}Number of \\ Non-bacilli\\ regions\end{tabular}} & \textbf{Training Accuracy} \\ \hline
\textbf{Random Forest} & 250                                                                              & 250                                                                                 & 0.983333          \\ \hline
\textbf{XGBoost}       & 250                                                                              & 250                                                                                 & 0.986667          \\ \hline
\textbf{SVM}           & 250                                                                              & 250                                                                                 & 0.976667          \\ \hline
\textbf{Ensemble}      & 750                                                                              & 750                                                                                 & 0.983660          \\ \hline
\end{tabular}
\end{table}

\begin{figure}[htp]
  \begin{center}
    \subfigure[Original RGB test image]{\label{va1}\includegraphics[scale=0.5]{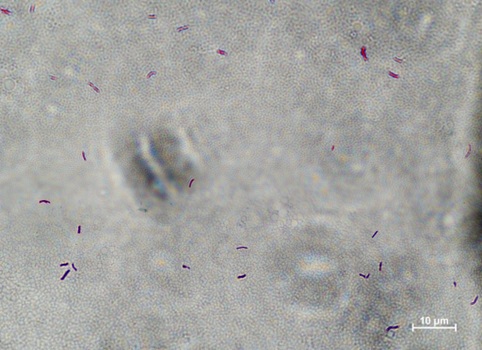}}
    \subfigure[Ground truth mask]{\label{va2}\includegraphics[scale=0.51]{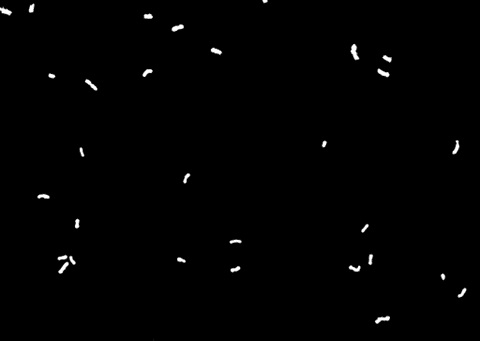}} 
    \subfigure[Segmentation - proposed model]{\label{va3}\includegraphics[scale=0.5]{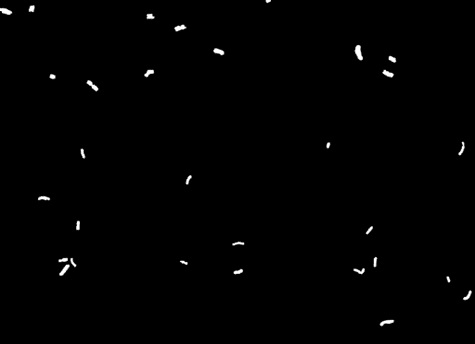}}
    \subfigure[Otsu segmentation described in \cite{panicker2018automatic}]{\label{va4}\includegraphics[scale=0.085]{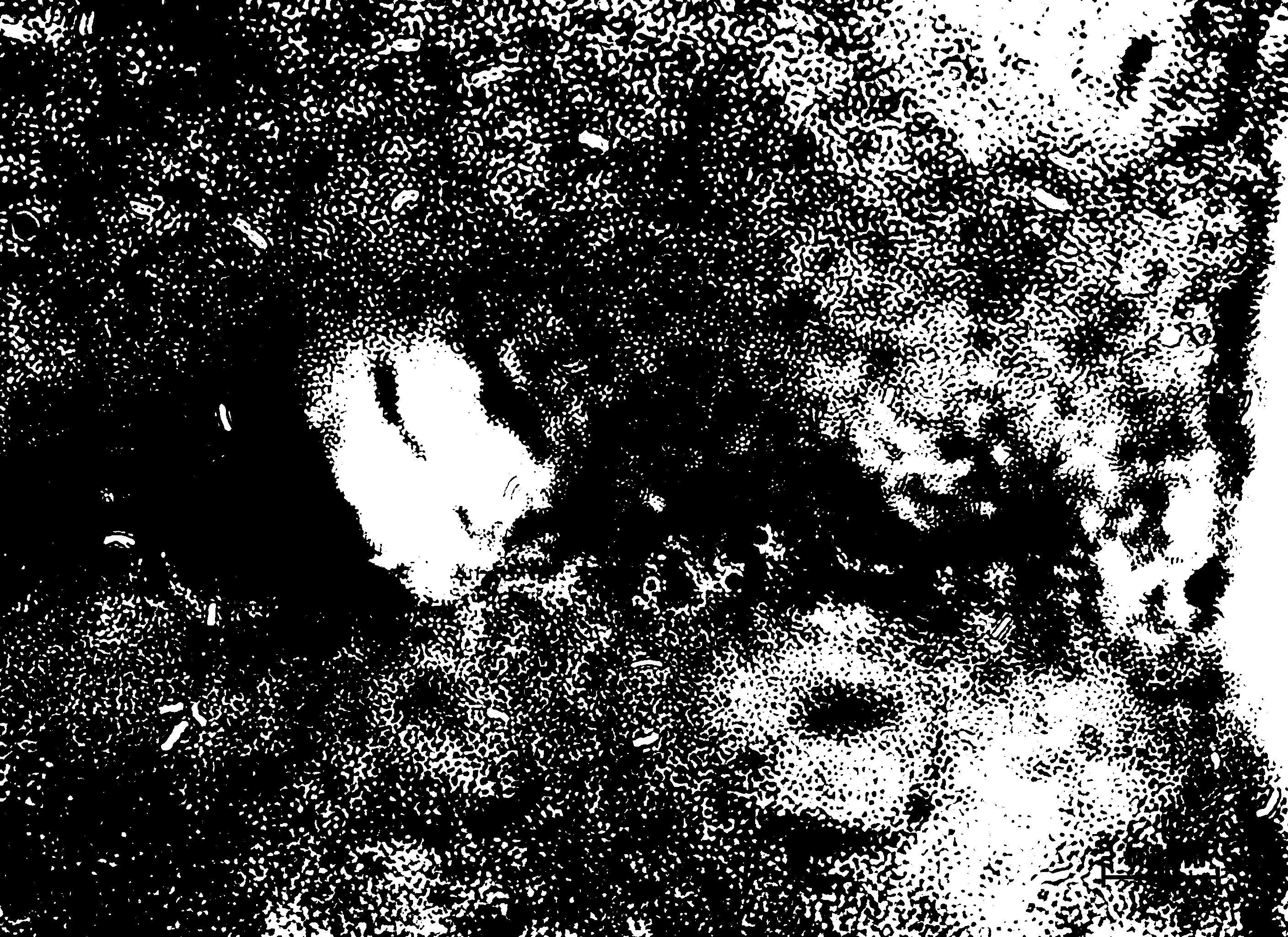}}
    \subfigure[Otsu segmentation described in \cite{mithra2023enhanced}]{\label{va5}\includegraphics[scale=0.5]{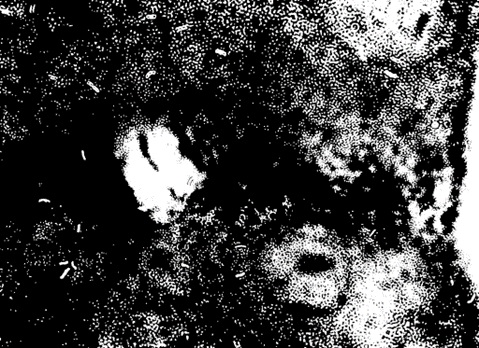}}
  \end{center}
  \caption{Segmentation results of Attention Residual U-Net and Otsu methods described in existing methods}
  \label{va}
\end{figure}

During testing of the segmentation model, similar to the methodology used in the training phase, the 20 test images were partitioned into patches of dimensions 256 $\times$ 256, resulting in a total of 1760 patches. The selection process of 20 test images, ensured randomness and equality across the three image sets. Specifically, six images were randomly chosen from the first slide, and seven images each were selected from the other two slides for testing. The patches underwent segmentation using the trained Attention Residual U-Net to obtain masks for each patch. Subsequently, the predicted masks of individual patches within an image are merged to generate a unified mask corresponding to each image in the test set. The segmentation performance of the Attention Residual U-Net, compared with the segmentation methods in \cite{panicker2018automatic} and \cite{mithra2023enhanced}, on DCA-CUSAT TB dataset is presented in Table \ref{seg_per1}.  Figure \ref{va} shows the segmentation results of Attention Residual U-Net on a sample test image from the DCA-CUSAT TB dataset. The comparative analysis in Table \ref{seg_per1} indicates that the proposed method significantly outperforms in segmentation methods described in \cite{panicker2018automatic} and \cite{mithra2023enhanced}. This superiority is further reinforced by the visual examination of the Figure \ref{va}.
\begin{table}[htp]
\caption{Performance of segmentation using existing methods and proposed method.}
\label{seg_per1}
\centering
\scalebox{0.7}

\begin{tabular}{|c|c|c|c|c|}
\hline
Dataset                       & \begin{tabular}[c]{@{}c@{}}Evaluation \\ metric\end{tabular} & \multicolumn{1}{l|}{\begin{tabular}[c]{@{}l@{}}Method \\ in \cite{panicker2018automatic}\end{tabular}} & \multicolumn{1}{l|}{\begin{tabular}[c]{@{}l@{}}Method \\ in \cite{mithra2023enhanced}\end{tabular}} & \multicolumn{1}{l|}{\textbf{\begin{tabular}[c]{@{}l@{}}Proposed \\ method\end{tabular}}} \\ \hline
\multirow{2}{*}{DCA-CUSAT TB} & Jaccard Index                                                  & 0.6834                                                                       & 0.7414                                                                       & \textbf{0.9360}                                                                          \\ \cline{2-5} 
                              & Dice coefficient                                               & 0.8119                                                                       & 0.8138                                                                       & \textbf{0.9670}                                                                          \\ \hline
\multirow{2}{*}{Costa}        & Jaccard Index                                                  & 0.4768                                                                       & 0.6361                                                                       & \textbf{0.9845}                                                                          \\ \cline{2-5} 
                              & Dice coefficient                                               & 0.6457                                                                       & 0.7776                                                                       & \textbf{0.9922}                                                                          \\ \hline
\multirow{2}{*}{ZNSM-iDB}     & Jaccard Index                                                  & 0.7328                                                                       & 0.7715                                                                       & \textbf{0.9767}                                                                          \\ \cline{2-5} 
                              & Dice coefficient                                               & 0.8457                                                                       & 0.8710                                                                       & \textbf{0.9882}                                                                          \\ \hline
\end{tabular}
\end{table}

\par 
For testing the performance of the proposed bacilli detection method after segmentation, contour analysis is performed to identify bacilli suspected regions on each mask corresponding to the images in the dataset. These regions are extracted from the images which serve as the Regions of Interest (RoIs). The RoIs are subsequently classified as either bacilli or non-bacilli using the trained ensemble classifier. The proposed method is compared with the existing methods\cite{panicker2018automatic} and \cite{mithra2023enhanced} on DCA CUSAT TB dataset and the comparative results are presented in Table \ref{pcr_ec}. The quantitative results indicating accuracy, precision, recall and F1 score, presented in Table \ref{pcr_ec} demonstrate that the proposed model exhibits superior performance compared to other existing methods for bacilli identification.

\begin{table}[htp]
\centering
\caption{Performance comparison of existing methods and proposed method on the DCA-CUSAT TB dataset}
\label{pcr_ec}
\begin{tabular}{|c|c|c|c|c|}
\hline
Method                                                                                  & Accuracy & Precision & Recall & F1-Score \\ \hline
\begin{tabular}[c]{@{}c@{}}Method in \cite{panicker2018automatic}\end{tabular} & 0.6773   & 0.6711    & 0.6    & 0.6335   \\ \hline
\begin{tabular}[c]{@{}c@{}}Method in \cite{mithra2023enhanced}\end{tabular}                & 0.7119   & 0.9       & 0.75   & 0.8181     \\ \hline
\begin{tabular}[c]{@{}c@{}}\textbf{Proposed} \textbf{Method}\end{tabular}                              & \textbf{0.9654}     &\textbf{ 0.9723}      &\textbf{ 0.9923}   & \textbf{0.9822  }   \\ \hline
\end{tabular}
\end{table}

\subsection{Performance Analysis of the Proposed Model Using Costa Dataset}

The proposed model is evaluated using the Costa dataset \cite{ya12-j913-22} for comparative analysis. From this dataset, a total of 90 images were randomly selected, representing both high and low background densities. Additionally, these images include varying bacilli densities, with some having a high density of bacilli and others a low density. Out of these, 72 images were utilized for the training phase of the proposed model, while the remaining 18 images were reserved for testing purposes. All images had a resolution of 2816 $\times$ 2048. 

\par
\begin{figure}[htp]
  \begin{center}
    \subfigure[Training accuracy]{\label{sam1}\includegraphics[scale=0.5]{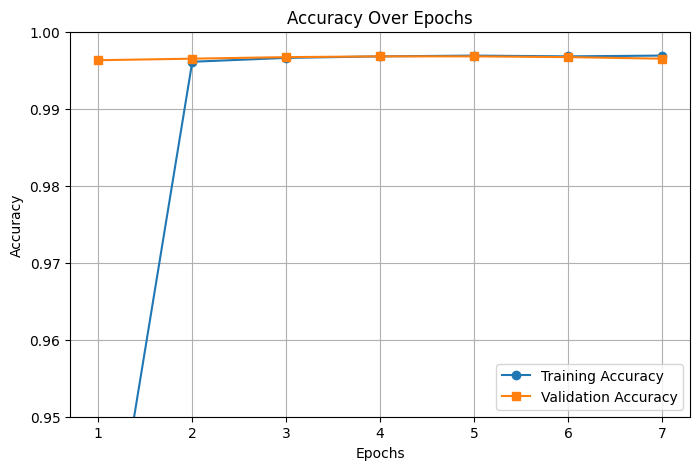}}
    \subfigure[Jaccard Index]{\label{sam2}\includegraphics[scale=0.5]{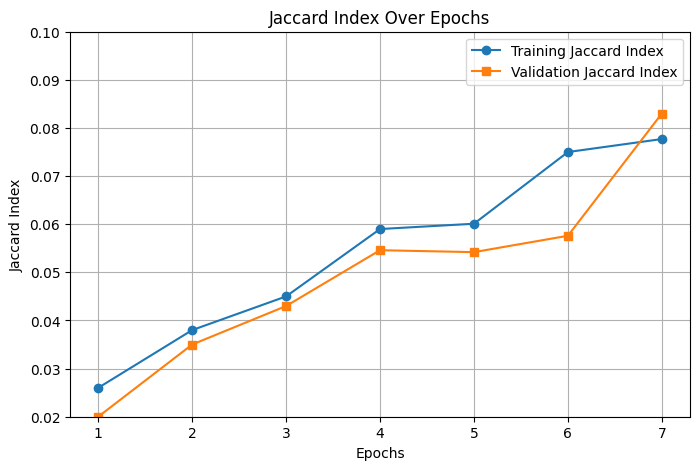}} 
    \subfigure[Loss]{\label{sam3}\includegraphics[scale=0.5]{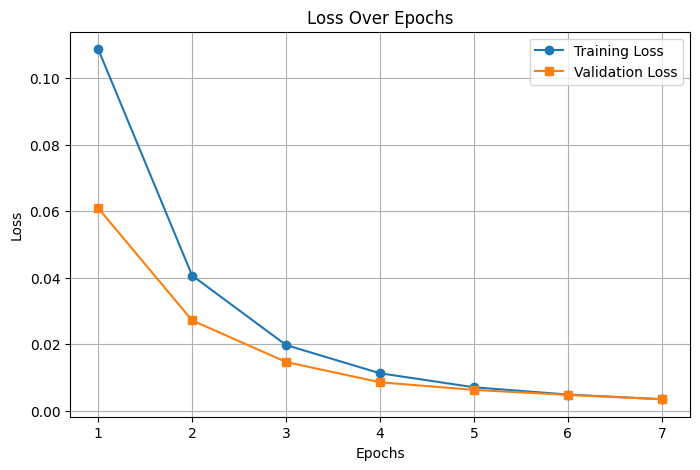}}
  \end{center}
  \caption{The progress of the Attention Residual U-Net training on Costa dataset.}
  \label{gr3}
\end{figure}

The training phase of the segmentation model utilizes a substantial dataset derived from 72 images. A total of 6336 patches of size 256 $\times$ 256 were produced from these images. These patches, along with the corresponding ground truth masks, were used to train the proposed Attention Residual U-Net model, which exhibited high accuracy. The accuracy during the training phase was 99.65\% and is graphically plotted in Figure \ref{gr3} along with jaccard index and loss.

\par 

To train the ensemble classifier, 750 segments containing bacilli and 750 segments containing non-bacilli regions were extracted. Initially, each classifier within the ensemble model underwent training using 250 bacilli regions and 250 non-bacilli regions. Following this, the ensemble model was trained using the combined dataset. The accuracy of each individual classifier, as well as the ensemble classifier on the Costa dataset, is presented in Table \ref{tr_acc_various}.

\begin{table}[htp]
\centering
\caption{Training accuracy of the proposed model on various datasets}
\label{tr_acc_various}
\begin{tabular}{|c|c|c|c|}
\hline
\textbf{Classifier}                                            & \textbf{DCA-CUSAT TB} & \textbf{Costa} & \textbf{ZNSM-iDB} \\ \hline
SVM                                                            & 0.976667     & 0.980198       & 0.970873          \\ \hline
Random Forest                                                  & 0.983333     & 0.990099       & 0.950980          \\ \hline
XGBoost                                                        & 0.986667     & 0.931372       & 0.981325          \\ \hline
\begin{tabular}[c]{@{}c@{}}Ensemble \\ Classifier\end{tabular} & 0.983660     & 0.980263       & 0.9836601         \\ \hline
\end{tabular}
\end{table}

In the testing phase of the segmentation model, the trained Attention Residual U-Net accurately segmented 1584 patches extracted from 18 test images. These segmented patches were then utilized to reconstruct the corresponding masks for each image. The segmentation performance of the proposed method along with the performance of existing methods in \cite{panicker2018automatic} and \cite{mithra2023enhanced}, on the Costa dataset is presented in Table \ref{seg_per1}.

\par 
The Regions of Interest (RoIs) extracted from original images, using the proposed segmentation model and contour analysis, are classified using the trained ensemble classifier. The performance analysis of the proposed technique in comparison with the existing methods on Costa dataset is shown in Table \ref{per_en_ex}.

\begin{table}[htp]
\centering
\caption{Performance comparison of existing methods and proposed method on the Costa dataset}
\label{per_en_ex}
\begin{tabular}{|c|c|c|c|c|}
\hline
\textbf{Method}                                                                                            & \textbf{Accuracy} & \textbf{Precision} & \textbf{Recall} & \textbf{F1-Score} \\ \hline
\begin{tabular}[c]{@{}c@{}}Method  in \cite{panicker2018automatic}\end{tabular} & 0.6               & 0.67               & 0.6             & 0.6330              \\ \hline
\begin{tabular}[c]{@{}c@{}}Method in \cite{mithra2023enhanced}\end{tabular}    & 0.73              & 0.8                & 0.758           & 0.7784              \\ \hline
\textbf{\begin{tabular}[c]{@{}c@{}}Proposed Method\end{tabular}}                                        & \textbf{0.9443}     & \textbf{0.9766}      & \textbf{0.9480}   & \textbf{0.9621}     \\ \hline
\end{tabular}
\end{table}

\par 
The superiority of the proposed model over other methodologies is evident from Table \ref{per_en_ex}, which showcases its exceptional performance on the Costa dataset as well.

\subsection{Performance Analysis of the Proposed Model Using ZNSM-iDB Dataset}

For performance analysis, the proposed method is applied on the ZNSM-iDB \cite{shah2017ziehl} dataset, which includes images of varying resolutions and backgrounds with different colors. 90 images, each with a resolution of 2592 $\times$ 1944, were randomly sampled from this dataset. Among these, 72 images were designated for the training set, while the remaining 18 images were allocated for testing purposes. 
\par
\begin{figure}[htp]
  \begin{center}
    \subfigure[Training accuracy]{\label{sam1}\includegraphics[scale=0.5]{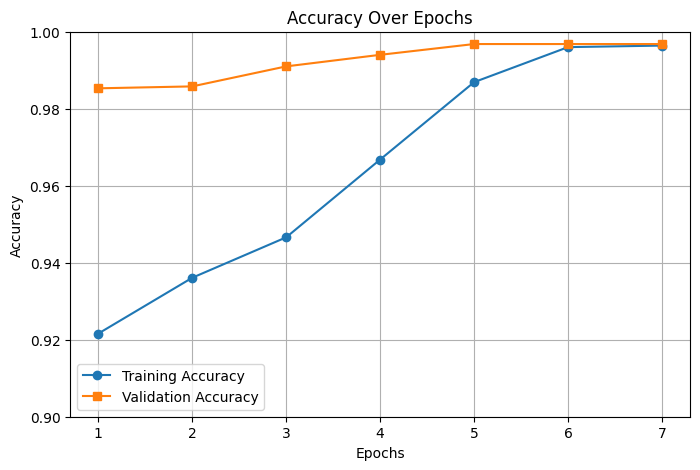}}
    \subfigure[Jaccard Index]{\label{sam2}\includegraphics[scale=0.5]{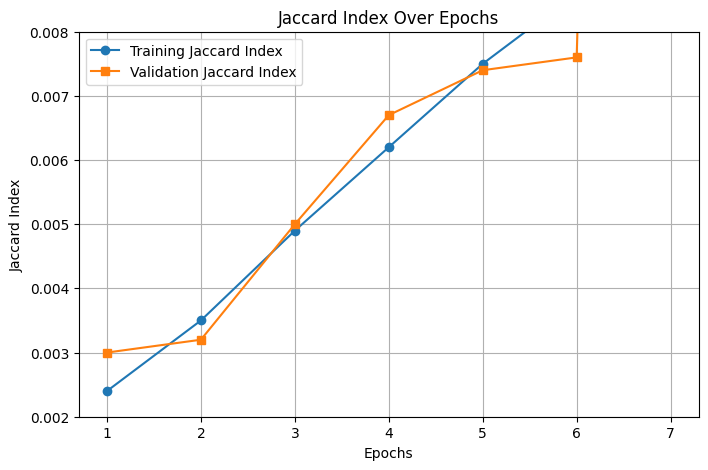}} 
    \subfigure[Loss]{\label{sam3}\includegraphics[scale=0.5]{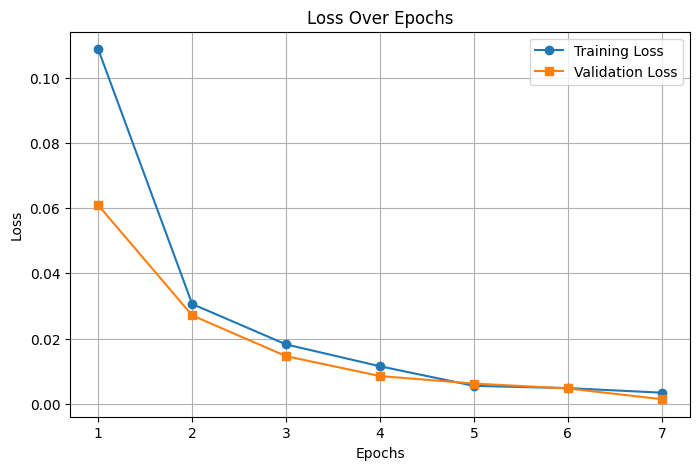}}
  \end{center}
  \caption{The progress of Attention Residual U-Net training on ZNSM-iDB dataset.}
  \label{gr2}
\end{figure}

The training phase of the segmentation model utilizes 5040 patches generated from 72 images, each sized at 256 $\times$ 256. The training patches, along with their corresponding masks, were utilized to train the Attention Residual U-Net model, achieving an accuracy of 99.68\%. The values of accuracy, jaccard index and loss during training are depicted in Figure \ref{gr2}.

\par
To train the ensemble classifier, similar to the methodologies applied to other datasets, both bacilli and non-bacilli regions were extracted from this dataset as well. 250 regions from each category were used to train each of the three individual classifiers. Subsequently, the ensemble classifier was trained using a combined dataset comprising 750 bacilli and 750 non-bacilli regions. The accuracy achieved during the training period is detailed in Table \ref{tr_acc_various}.

\par

After the training procedures, 1260 test patches have been segmented using the Attention Residual U-Net model. Each patch, sized 256 $\times$ 256 is formed by splitting the 18 test images in the test set of ZNSM-iDB dataset. 
The performance of the proposed segmentation model, as well as the existing segmentation methods described in \cite{panicker2018automatic} and \cite{mithra2023enhanced}, on the ZNSM-iDB is provided in Table \ref{seg_per1}.
\par
Finally the tuberculosis bacillus are identified from each test set image by using the proposed methodology. The performance of the proposed methodology on ZNSM-iDB is quantitatively assessed and compared against the existing techniques  in \cite{panicker2018automatic} and \cite{mithra2023enhanced}. The comparative results are given in Table \ref{znaccuracy}. It clearly indicates that the proposed method excels compared to existing methods in terms of accuracy, precision, recall, and F1 score.

\begin{table}[htp]
\centering
\caption{Performance of various identification methods on the ZNSM-iDB dataset.}
\label{znaccuracy}
\begin{tabular}{|c|c|c|c|c|}
\hline
\textbf{Method}                                                                                            & \textbf{Accuracy} & \textbf{Precision} & \textbf{Recall} & \textbf{F1-Score} \\ \hline
\begin{tabular}[c]{@{}c@{}}Method in \cite{panicker2018automatic}\end{tabular} & 0.63              & 0.6                & 0.66            & 0.6285              \\ \hline
\begin{tabular}[c]{@{}c@{}}Method in \cite{mithra2023enhanced}\end{tabular}    & 0.74              & 0.7                & 0.765           & 0.7310             \\ \hline
\textbf{\begin{tabular}[c]{@{}c@{}}Proposed Method\end{tabular}}                                        & \textbf{0.9107}     & \textbf{0.9112}      & \textbf{0.9495}   & \textbf{0.9303}     \\ \hline
\end{tabular}
\end{table}

\par
Table \ref{seg_per2} presents the comprehensive information regarding the images and patch divisions employed for training and testing of the proposed segmentation model. It encompasses image resolutions, the quantity of training and testing images, as well as the number of training and testing patches.

\begin{table}[htp]
\caption{Details of images and patch divisions for training and testing of the Attention Residual U-Net for segmentation on different datasets.}
\label{seg_per2}
\centering
\scalebox{0.7}{
\begin{tabular}{|c|c|c|c|c|c|}
\hline
\textbf{Database}    & \textbf{\begin{tabular}[c]{@{}c@{}}Image \\ Resolution\end{tabular}} & \textbf{\begin{tabular}[c]{@{}c@{}}Number of \\ Training images\end{tabular}} & \textbf{\begin{tabular}[c]{@{}c@{}}Number of \\ patches \\ (training)\end{tabular}} & \textbf{\begin{tabular}[c]{@{}c@{}}Test \\ images\end{tabular}} & \textbf{\begin{tabular}[c]{@{}c@{}}Number of \\ patches \\ (testing)\end{tabular}}  \\ \hline
\textbf{DCA-CUSAT TB Dataset} & 2880 $\times$ 2048                                                          & 81                                                                            & 7128                                                                                & 20                                                              & 1760                                                                                                                             \\ \hline
\textbf{Costa}       & 2816 $\times$ 2048                                                          & 72                                                                            & 6336                                                                                & 18                                                              & 1584                                                                                                                             \\ \hline
\textbf{ZNSM-iDB}    & 2560 $\times$ 1792                                                          & 72                                                                            & 5040                                                                                & 18                                                              & 1260                                                                                                                             \\ \hline
\end{tabular}}
\end{table}

\par

\par 
In  \cite{panicker2018automatic} and \cite{mithra2023enhanced}, the two customized variants of Otsu algorithm were utilized for image binarization, which demands lower computational resources compared to the proposed segmentation model. However, these approaches yielded significantly lower values for both the jaccard index and dice coefficient as given in Table \ref{seg_per1} which indicates inferior segmentation outcomes. Moreover the visual inspection of Figure \ref{va} clearly indicates that our proposed segmentation model exceeds the customized Otsu segmentation methods from  \cite{panicker2018automatic} and \cite{mithra2023enhanced}, showing a close match with the ground truth mask. 

\par
The introduction of an efficient Attention Residual U-Net in the segmentation stage necessitates a slightly extended training time.However, the time needed to train the ensemble classifier in the classification stage is comparable to that of existing methods. Performance analyses conducted with the proposed methodology and various existing methods across different datasets  clearly demonstrates its superiority over existing methods in accurately identifying bacilli regions within microscopic sputum smear images.

\section{Conclusion}
This paper introduces an efficient technique for detecting tuberculosis bacilli from bright field microscopic sputum smear images using Attention Residual U-Net in segmentation stage and an ensemble classifier combining Random Forest, XGBoost, and SVM classifiers in classification stage. A new database named 'DCA-CUSAT Bright Field Microscopic Sputum Smear TB Dataset' has been constructed consisting of 101 microscopic sputum smear images, to conduct the experiments. Furthermore, the proposed methodology is evaluated against existing methods by applying them on the Costa dataset, ZNSM-iDB dataset, and DCA-CUSAT TB dataset. The increased automation, good segmentation performance, and accuracy in classification indicate the potential of the proposed approach to enhance tuberculosis detection and treatment. Future studies could concentrate on enlarging the dataset and refining the model to further improve its capabilities. Moreover, investigating techniques to count bacilli within clusters would enhance diagnostic precision and provide deeper insights into disease progression.

\begin{acknowledgements}
We extend our sincere thanks to Dr. Sarath G Rao, District TB officer and technical staff of District TB Hospital, Ernakulam, Kerala, India for  providing ZN-stained sputum smear samples necessary for capturing microscopic images, and providing ground truth information which were crucial for conducting our research. 
\end{acknowledgements}

%\bibliography{tb-bibfile}


\begin{thebibliography}{10}
\providecommand{\url}[1]{{#1}}
\providecommand{\urlprefix}{URL }
\expandafter\ifx\csname urlstyle\endcsname\relax
  \providecommand{\doi}[1]{DOI~\discretionary{}{}{}#1}\else
  \providecommand{\doi}{DOI~\discretionary{}{}{}\begingroup
  \urlstyle{rm}\Url}\fi

\bibitem{tbr2022}
India tb report 2022 coming together to end tb altogether pp. 1--145 (2022)

\bibitem{alom2018recurrent}
Alom, M.Z., Hasan, M., Yakopcic, C., Taha, T.M., Asari, V.K.: Recurrent
  residual convolutional neural network based on u-net (r2u-net) for medical
  image segmentation.
\newblock arXiv preprint arXiv:1802.06955  (2018)

\bibitem{ayas2014random}
Ayas, S., Ekinci, M.: Random forest-based tuberculosis bacteria classification
  in images of zn-stained sputum smear samples.
\newblock Signal, Image and Video Processing \textbf{8}, 49--61 (2014)

\bibitem{bankhead2017qupath}
Bankhead, P., Loughrey, M.B., Fern{\'a}ndez, J.A., Dombrowski, Y., McArt, D.G.,
  Dunne, P.D., McQuaid, S., Gray, R.T., Murray, L.J., Coleman, H.G., et~al.:
  Qupath: Open source software for digital pathology image analysis.
\newblock Scientific reports \textbf{7}(1), 1--7 (2017)

\bibitem{breiman2001random}
Breiman, L.: Random forests.
\newblock Machine learning \textbf{45}, 5--32 (2001)

\bibitem{chen2024cavmamba}
Chen, Q., Xu, Z., Fang, X.: Cavmamba: convolution-augmented vmamba for medical
  image segmentation.
\newblock The Visual Computer pp. 1--18 (2024)

\bibitem{chen2016xgboost}
Chen, T., Guestrin, C.: Xgboost: A scalable tree boosting system.
\newblock In: Proceedings of the 22nd acm sigkdd international conference on
  knowledge discovery and data mining, pp. 785--794 (2016)

\bibitem{costa2008automatic}
Costa, M.G., Costa~Filho, C.F., Sena, J.F., Salem, J., de~Lima, M.O.: Automatic
  identification of mycobacterium tuberculosis with conventional light
  microscopy.
\newblock In: 2008 30th Annual International Conference of the IEEE Engineering
  in Medicine and Biology Society, pp. 382--385. IEEE (2008)

\bibitem{costa2015automatic}
Costa~Filho, C.F.F., Levy, P.C., Xavier, C.d.M., Fujimoto, L.B.M., Costa,
  M.G.F.: Automatic identification of tuberculosis mycobacterium.
\newblock Research on biomedical engineering \textbf{31}, 33--43 (2015)

\bibitem{costafilho2012mycobacterium}
CostaFilho, C.F., Levy, P.C., Xavier, C.M., Costa, M.G., Fujimoto, L.B., Salem,
  J.: Mycobacterium tuberculosis recognition with conventional microscopy.
\newblock In: 2012 Annual International Conference of the IEEE Engineering in
  Medicine and Biology Society, pp. 6263--6268. IEEE (2012)

\bibitem{dinesh2019tuberculosis}
Dinesh Jackson~Samuel, R., Rajesh~Kanna, B.: Tuberculosis (tb) detection system
  using deep neural networks.
\newblock Neural Computing and Applications \textbf{31}, 1533--1545 (2019)

\bibitem{el2019identification}
El-Melegy, M., Mohamed, D., ElMelegy, T., Abdelrahman, M.: Identification of
  tuberculosis bacilli in zn-stained sputum smear images: A deep learning
  approach.
\newblock In: Proceedings of the IEEE/CVF Conference on Computer Vision and
  Pattern Recognition Workshops, pp. 0--0 (2019)

\bibitem{fandriyanto2021detecting}
Fandriyanto, A., Rajab, T.E., et~al.: Detecting and counting tuberculosis
  bacilli in a microscopic image of patient sputum using gaussian mixture
  model.
\newblock In: 2021 International Conference on Artificial Intelligence and
  Computer Science Technology (ICAICST), pp. 84--89. IEEE (2021)

\bibitem{greeshma2023identification}
Greeshma, K., Vishnukumar, S.: Identification of tuberculosis bacilli from
  bright field microscopic sputum smear images using u-net and random forest
  classification algorithm.
\newblock In: 2023 International Conference on Advances in Intelligent
  Computing and Applications (AICAPS), pp. 1--4. IEEE (2023)

\bibitem{hearst1998support}
Hearst, M.A., Dumais, S.T., Osuna, E., Platt, J., Scholkopf, B.: Support vector
  machines.
\newblock IEEE Intelligent Systems and their applications \textbf{13}(4),
  18--28 (1998)

\bibitem{kant2018towards}
Kant, S., Srivastava, M.M.: Towards automated tuberculosis detection using deep
  learning.
\newblock In: 2018 IEEE Symposium Series on Computational Intelligence (SSCI),
  pp. 1250--1253. IEEE (2018)

\bibitem{khutlang2009classification}
Khutlang, R., Krishnan, S., Dendere, R., Whitelaw, A., Veropoulos, K.,
  Learmonth, G., Douglas, T.S.: Classification of mycobacterium tuberculosis in
  images of zn-stained sputum smears.
\newblock IEEE transactions on information technology in biomedicine
  \textbf{14}(4), 949--957 (2009)

\bibitem{khutlang2010automated}
Khutlang, R., Krishnan, S., Whitelaw, A., Douglas, T.S.: Automated detection of
  tuberculosis in ziehl-neelsen-stained sputum smears using two one-class
  classifiers.
\newblock Journal of microscopy \textbf{237}(1), 96--102 (2010)

\bibitem{liu2021nhbs}
Liu, R., Liu, M., Sheng, B., Li, H., Li, P., Song, H., Zhang, P., Jiang, L.,
  Shen, D.: Nhbs-net: A feature fusion attention network for ultrasound
  neonatal hip bone segmentation.
\newblock IEEE Transactions on Medical Imaging \textbf{40}(12), 3446--3458
  (2021)

\bibitem{lopez2017automatic}
L{\'o}pez, Y.P., Costa~Filho, C., Aguilera, L., Costa, M.: Automatic
  classification of light field smear microscopy patches using convolutional
  neural networks for identifying mycobacterium tuberculosis.
\newblock In: 2017 CHILEAN conference on electrical, electronics engineering,
  information and communication technologies (CHILECON), pp. 1--5. IEEE (2017)

\bibitem{ya12-j913-22}
Marly G~F, C., Cicero F~F, C.F., Luciana~B, F., Pamela~C, L., Clahildek~M, X.:
  Tbimages: a smear microscopy image dataset to support the development of
  automated bacilli detection in the diagnosis of tuberculosis (2022)

\bibitem{mithra2023enhanced}
Mithra, K.: Enhanced fuzzy gaussian networks for sputum image based
  mycobacterium detection.
\newblock The Imaging Science Journal \textbf{71}(4), 313--322 (2023)

\bibitem{nazir2020off}
Nazir, A., Cheema, M.N., Sheng, B., Li, H., Li, P., Yang, P., Jung, Y., Qin,
  J., Kim, J., Feng, D.D.: Off-enet: An optimally fused fully end-to-end
  network for automatic dense volumetric 3d intracranial blood vessels
  segmentation.
\newblock IEEE Transactions on Image Processing \textbf{29}, 7192--7202 (2020)

\bibitem{nazir2021ecsu}
Nazir, A., Cheema, M.N., Sheng, B., Li, P., Li, H., Xue, G., Qin, J., Kim, J.,
  Feng, D.D.: Ecsu-net: an embedded clustering sliced u-net coupled with fusing
  strategy for efficient intervertebral disc segmentation and classification.
\newblock IEEE Transactions on Image Processing \textbf{31}, 880--893 (2021)

\bibitem{oktay2018attention}
Oktay, O., Schlemper, J., Folgoc, L.L., Lee, M., Heinrich, M., Misawa, K.,
  Mori, K., McDonagh, S., Hammerla, N.Y., Kainz, B., et~al.: Attention u-net:
  Learning where to look for the pancreas.
\newblock arXiv preprint arXiv:1804.03999  (2018)

\bibitem{whotbr2022}
Organization, W.H.: Global tuberculosis report 2022 pp. 1--68 (2022)

\bibitem{whotbr2023}
Organization, W.H.: Global tuberculosis report 2023 pp. 1--75 (2023)

\bibitem{panicker2018automatic}
Panicker, R.O., Kalmady, K.S., Rajan, J., Sabu, M.: Automatic detection of
  tuberculosis bacilli from microscopic sputum smear images using deep learning
  methods.
\newblock Biocybernetics and Biomedical Engineering \textbf{38}(3), 691--699
  (2018)

\bibitem{panicker2021lightweight}
Panicker, R.O., Pawan, S., Rajan, J., Sabu, M.: A lightweight convolutional
  neural network model for tuberculosis bacilli detection from microscopic
  sputum smear images.
\newblock Machine learning for healthcare applications pp. 343--351 (2021)

\bibitem{panicker2022automatic}
Panicker, R.O., Sabu, M.: Automatic detection of tuberculosis bacilli from
  conventional sputum smear microscopic images using densely connected
  convolutional networks.
\newblock SN Computer Science \textbf{3}(4), 263 (2022)

\bibitem{priya2015separation}
Priya, E., Srinivasan, S.: Separation of overlapping bacilli in microscopic
  digital tb images.
\newblock Biocybernetics and Biomedical Engineering \textbf{35}(2), 87--99
  (2015)

\bibitem{sadaphal2008image}
Sadaphal, P., Rao, J., Comstock, G., Beg, M.: Image processing techniques for
  identifying mycobacterium tuberculosis in ziehl-neelsen stains.
\newblock The International Journal of Tuberculosis and Lung Disease
  \textbf{12}(5), 579--582 (2008)

\bibitem{serrao2020automatic}
Serr{\~a}o, M., Costa, M.G.F., Fujimoto, L., Ogusku, M.M., Costa~Filho, C.F.F.:
  Automatic bacillus detection in light field microscopy images using
  convolutional neural networks and mosaic imaging approach.
\newblock In: 2020 42nd Annual International Conference of the IEEE Engineering
  in Medicine \& Biology Society (EMBC), pp. 1903--1906. IEEE (2020)

\bibitem{shah2017ziehl}
Shah, M.I., Mishra, S., Yadav, V.K., Chauhan, A., Sarkar, M., Sharma, S.K.,
  Rout, C.: Ziehl--neelsen sputum smear microscopy image database: a resource
  to facilitate automated bacilli detection for tuberculosis diagnosis.
\newblock Journal of Medical Imaging \textbf{4}(2), 027503--027503 (2017)

\bibitem{sotaquira2009detection}
Sotaquira, M., Rueda, L., Narvaez, R.: Detection and quantification of bacilli
  and clusters present in sputum smear samples: a novel algorithm for pulmonary
  tuberculosis diagnosis.
\newblock In: 2009 international conference on digital image processing, pp.
  117--121. IEEE (2009)

\bibitem{steingart2006fluorescence}
Steingart, K.R., Henry, M., Ng, V., Hopewell, P.C., Ramsay, A., Cunningham, J.,
  Urbanczik, R., Perkins, M., Aziz, M.A., Pai, M.: Fluorescence versus
  conventional sputum smear microscopy for tuberculosis: a systematic review.
\newblock The Lancet infectious diseases \textbf{6}(9), 570--581 (2006)

\bibitem{yang2020cnn}
Yang, M., Nurzynska, K., Walts, A.E., Gertych, A.: A cnn-based active learning
  framework to identify mycobacteria in digitized ziehl-neelsen stained human
  tissues.
\newblock Computerized Medical Imaging and Graphics \textbf{84}, 101752 (2020)

\bibitem{zhai2010automatic}
Zhai, Y., Liu, Y., Zhou, D., Liu, S.: Automatic identification of mycobacterium
  tuberculosis from zn-stained sputum smear: Algorithm and system design.
\newblock In: 2010 IEEE international conference on robotics and biomimetics,
  pp. 41--46. IEEE (2010)

\bibitem{zhao2024mfadu}
Zhao, Y., Zhang, G., Li, K., Zhu, Z., Li, X., Zhang, Y., Fan, Z.: Mfadu-net: an
  enhanced doubleu-net with multi-level feature fusion and atrous decoder for
  medical image segmentation.
\newblock The Visual Computer pp. 1--11 (2024)

\end{thebibliography}

%\begin{thebibliography}{}
% etc
%\end{thebibliography}

\end{document}